\documentclass[twoside]{dis04}
\def\runauthor{P J Bussey}
\def\shorttitle{Charm and beauty production at CDF}
\begin{document}

\title{CHARM AND BEAUTY PRODUCTION AT CDF}

\author{PETER J. BUSSEY\\ for the CDF Collaboration}

\address{Department of Physics and Astronomy\\
University of Glasgow, Glasgow G12 8QQ, UK\\
E-mail: p.bussey@physics.gla.ac.uk }

\maketitle

\abstracts{A summary is presented of recent results from the CDF collaboration
on the production of charm and beauty hadrons.
}

\section{Introduction}
The Tevatron at Fermilab is at present the world's highest energy
particle accelerator.  After extensive refurbishments, data taking for
Run II started in 2001, and by the end of 2003 over 200 pb$^{-1}$ of
usable integrated luminosity had been accumulated by each Collider experiment,
double the total for Run I.

In preparation for Run II, the CDF detector also underwent an
extensive upgrade.  Of special importance for the results presented
here are the new Silicon Vertex Tracker, and a number of significant
improvements to the trigger system.  The Extremely Fast Tracker allows
tracks in the central tracking system to be identified in the
first-level trigger.  In the second-level trigger, these tracks can be
matched with tracking information from the SVT so as to identify
events containing a displaced secondary vertex, by identifying tracks
with a finite impact parameter relative to the beam line.  The events
are finally recorded after verification by the third-level trigger.

In this talk, recent measurements and analyses of charm
and beauty states using the CDF detector are presented. The emphasis
here is on production mechanisms; studies of the decay and lifetime properties
of such particles are given in other contributions to these Proceedings.

\section{Theoretical framework}
Quantum Chromo-Dynamics is currently our most basic theory to
describe the composition and production of hadrons.  The study of
particles containing $c$ and $b$ quarks provides important
perspectives on our understanding of the application of this theory,
because the heavy masses of these quarks allow perturbative
calculations to be employed in ways that differ from the
case of light partons.

The principal lowest-order QCD processes that contribute to the
production of heavy quarks $Q$ in $p\bar p$ collisions are the $q\bar q
\to Q\bar Q$ and $gg \to Q\bar Q$, processes.  When these so-termed 
{\it flavour creation\/} processes take place as hard scattering of
the initial-state quarks or gluons in the (anti-)proton, two heavy
quarks are produced at high transverse momentum $p_T$, and
subsequently fragment into hadrons.

There is a further possibility that an initial-state gluon in the
proton may split into a $Q\bar Q$ pair, and one of these may scatter at
high $p_T$ off a parton in the other incoming beam particle.  This
splitting may be calculated within perturbative QCD. Sometimes the
unscattered heavy quark is formed at high $p_T$, but most often it
will be at low $p_T$ and will not be detected.  In this case the heavy
quark pair may be regarded as part of the parton structure of the
proton.  These are known as {\it flavour excitation\/} processes.  A
third type of process involves the production of a $Q\bar Q$ pair via
the splitting of a gluon radiated within the fragmentation products of a
final-state high-$p_T$ parton. These may be referred to as {\it
fragmentation\/} heavy quark pairs.

Many QCD calculations have been performed at LO and NLO to evaluate
the various heavy quark production cross sections.  Such cross
sections normally need to be convoluted with a phenomenological
fragmentation function to evaluate the cross section for producing a
given charm or beauty particle at a given $p_T$ value.  At LO, the
familiar leading-log Monte Carlo programs PYTHIA and HERWIG have long
enabled the final-state hadronisation to be performed as part of the
event generation.  The recently-developed MC@NLO  program (S. Frixione et
al., see these Proceedings) does this also in
next-to-leading order parton calculations.

\begin{figure}[t]
\hspace*{1mm}
\epsfig{file=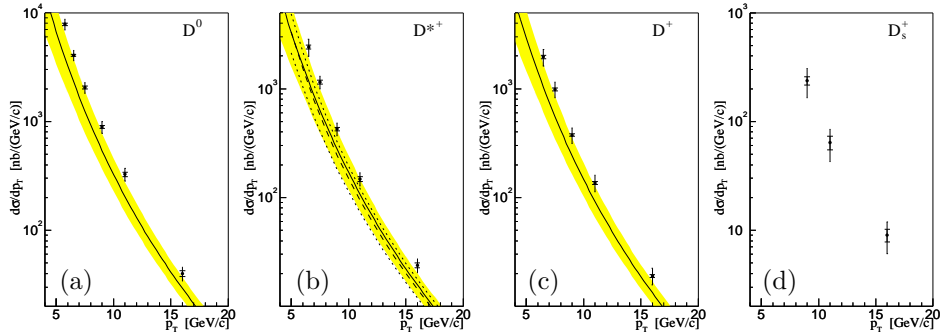,width=12.5cm}\\[-10mm]
\raisebox{0mm}{\hspace*{9mm}(a)\hspace*{26.9mm}(b)
\hspace{25mm}(c)\hspace{26.9mm}(d)}\\[3mm]
\caption{CDF inclusive cross sections for $D^0$, $D^{*+}$, $D^+$ and
$D_s^+$ as a function of transverse momentum $p_T$. Comparison is made
with the FONLL calculation of Cacciari and 
Nason \protect\cite{cacciari1} (a)--(c) and a similar
NLO calculation from Kniehl et al (priv.\ comm.) (b). 
No calculation was available for the $D_s$ case (d).} 
\end{figure}

\section{Open charm production}

The first published result from CDF II \cite{cdf1} consisted of a
precision measurement of the mass difference between the $D^+$ and
$D_s^+$ mesons, both observed in their decays to
$\phi\pi^+$. (Throughout this article, the mention of a state will
imply also the corresponding charge conjugate state.) Precision
was enhanced by a careful treatment of the material through which the
particles were tracked in the detector.  The measured value of $\Delta
m = 99.41
\pm 0.38 \pm 0.21$ MeV/c$^2$ improves on the previous PDG value.

Using just 6 pb$^{-1}$ of integrated luminosity, CDF have published
the first inclusive charm cross sections obtained at the
Tevatron \cite{cdf2}. The results are shown in fig.\ 1, compared with
the predictions from two NLO calculations. The FONLL model uses a
fixed-order NLO calculation employing next-to-leading logarithms and a
phenomenological $D$ fragmentation function obtained from ALEPH data,
with variable number of flavours.  The cross sections refer to
centrally produced $D$ mesons with rapidity in the range $[-1,
+1]$. The $p_T$ distribution is accurately described in shape by the
calculations, whose magnitude lies slightly lower than the data,
though still just about within the error band due to the renormalisation
and factorisation scale uncertainties.  Further uncertainties from the
mass of the charm quark are considered to be not larger than this.
The Kniehl calculation uses a different fragmentation scheme and mass
treatment from that of Cacciari et al. Overall, the agreement with
theory appears satisfactory.

\begin{figure}[!t]
\hspace*{4mm}
\epsfig{file=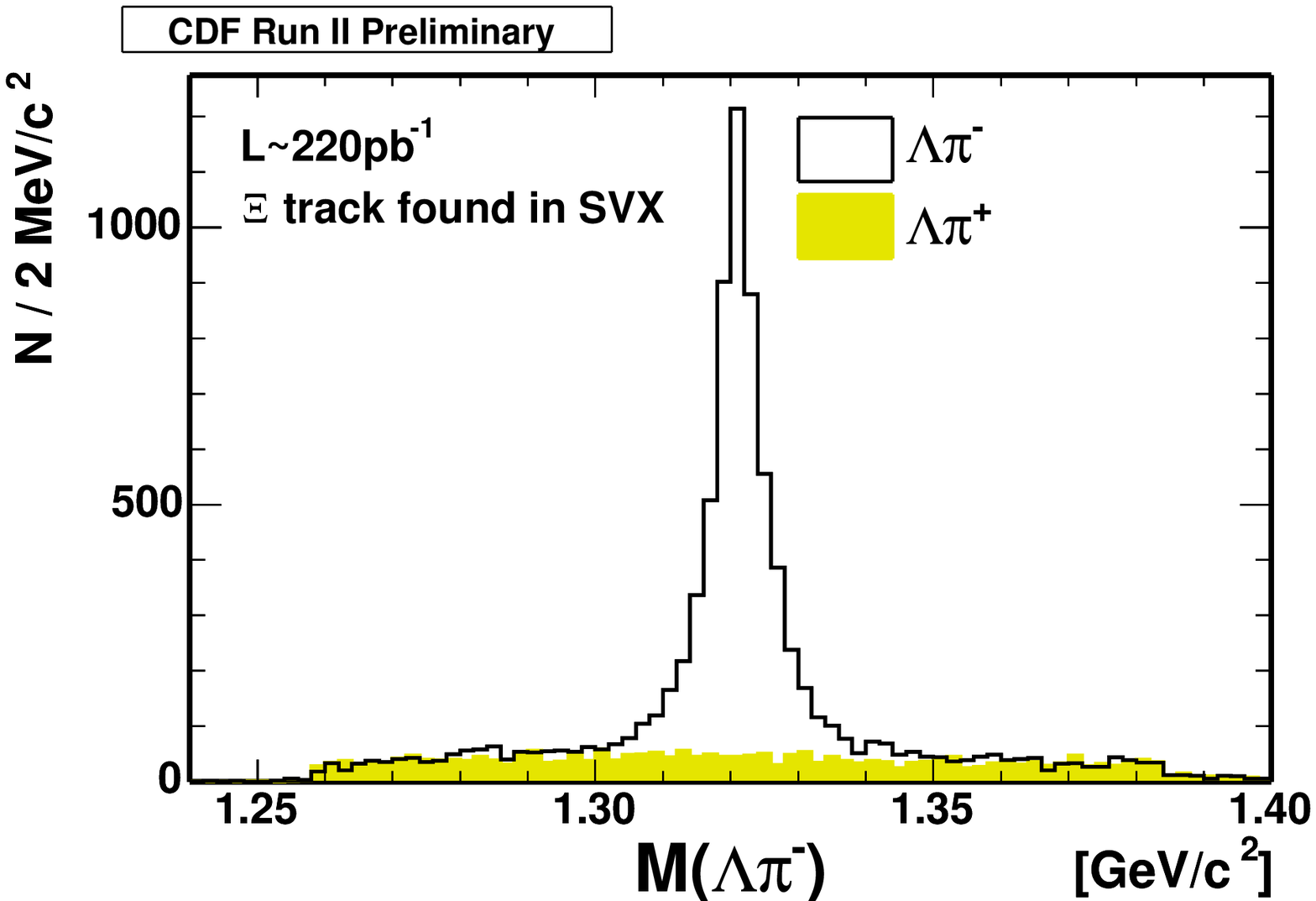,width=5.5cm}\hspace*{8mm}
\epsfig{file=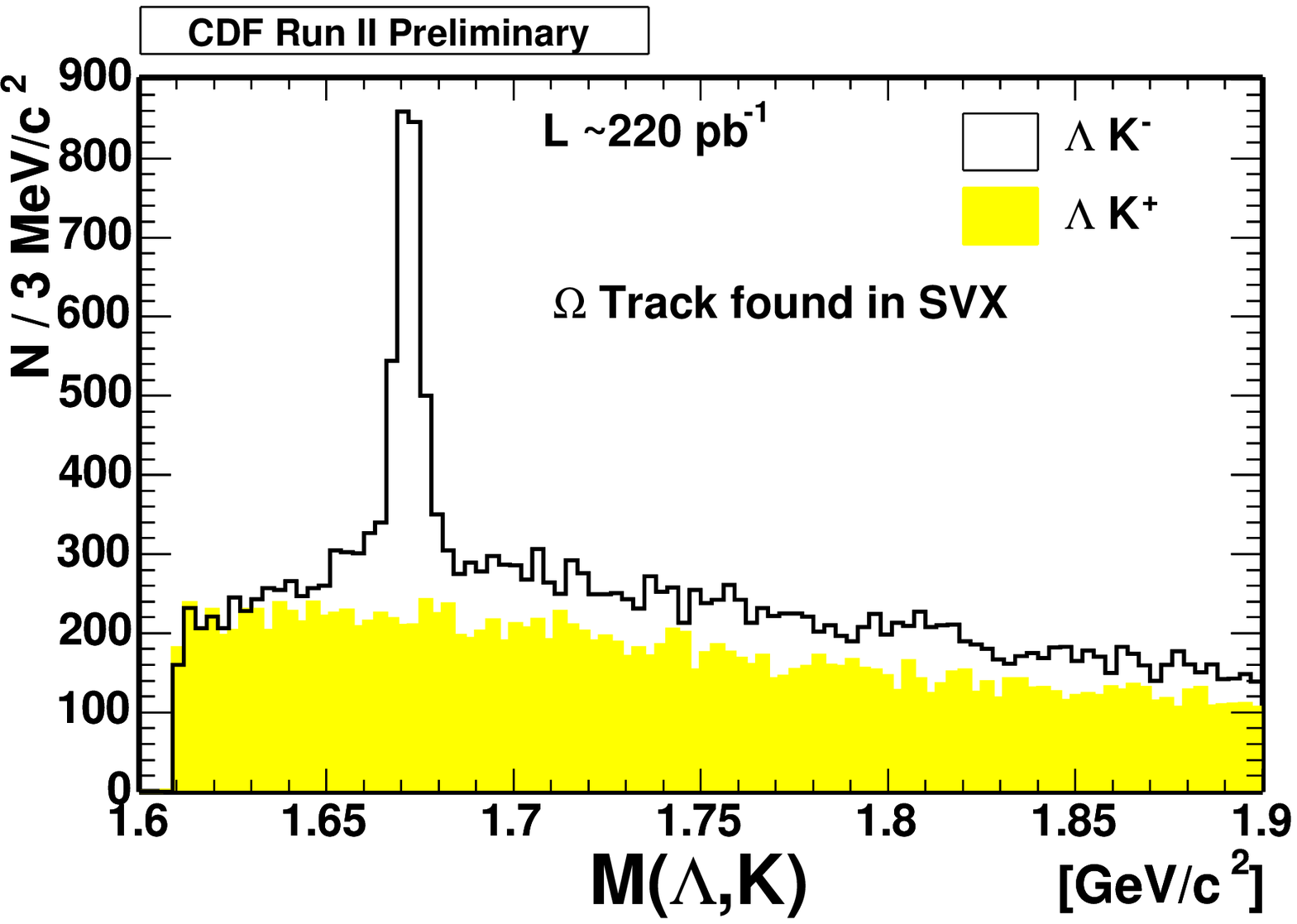,width=5.5cm}
\\[-4mm]
\hspace*{1mm}\raisebox{-1mm}{\hspace*{2mm}(a)\hspace*{60mm}(b)}\\[2mm]
\hspace*{6mm}
\epsfig{file=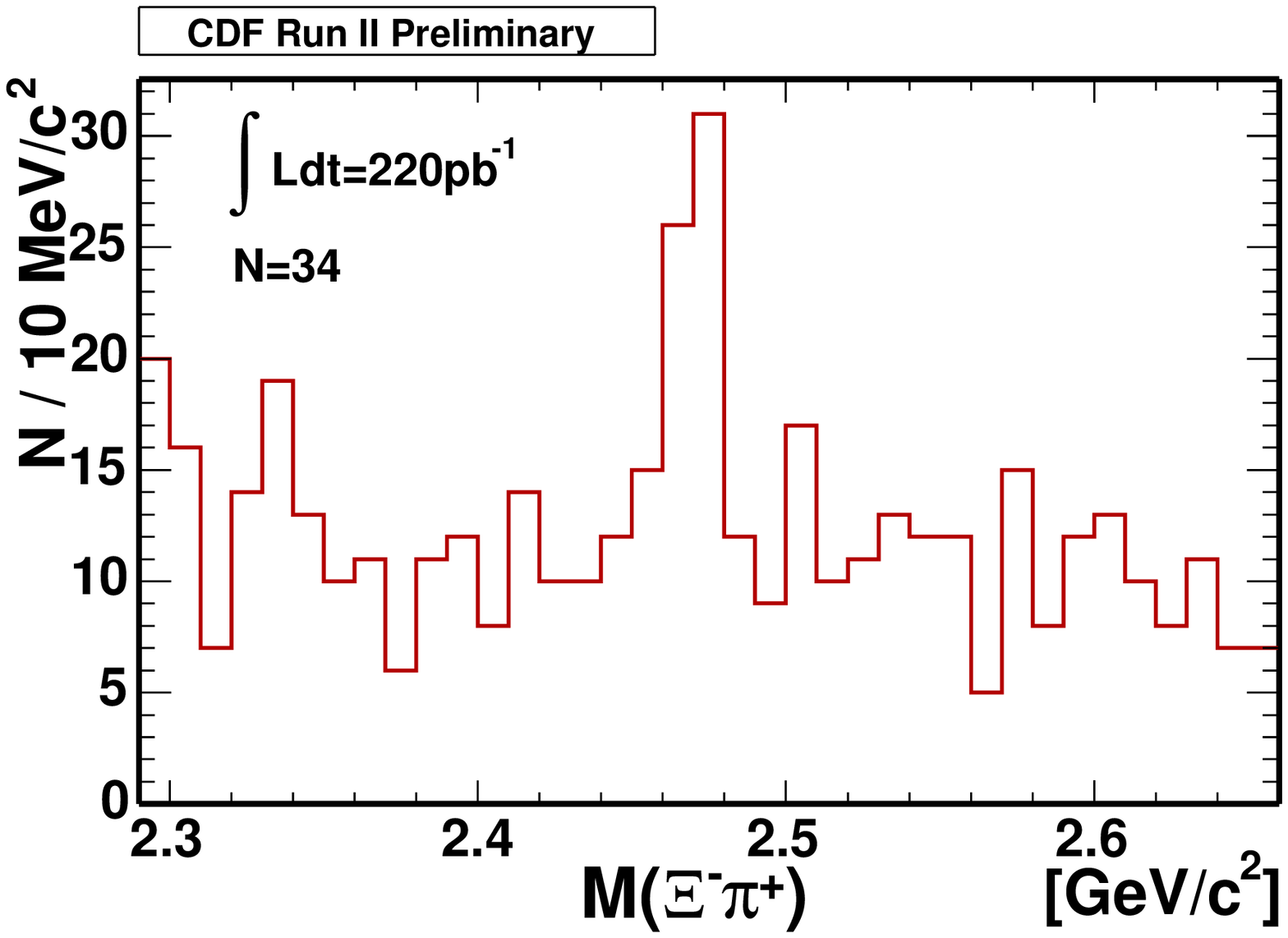,width=5.5cm}\hspace*{5mm}
\epsfig{file=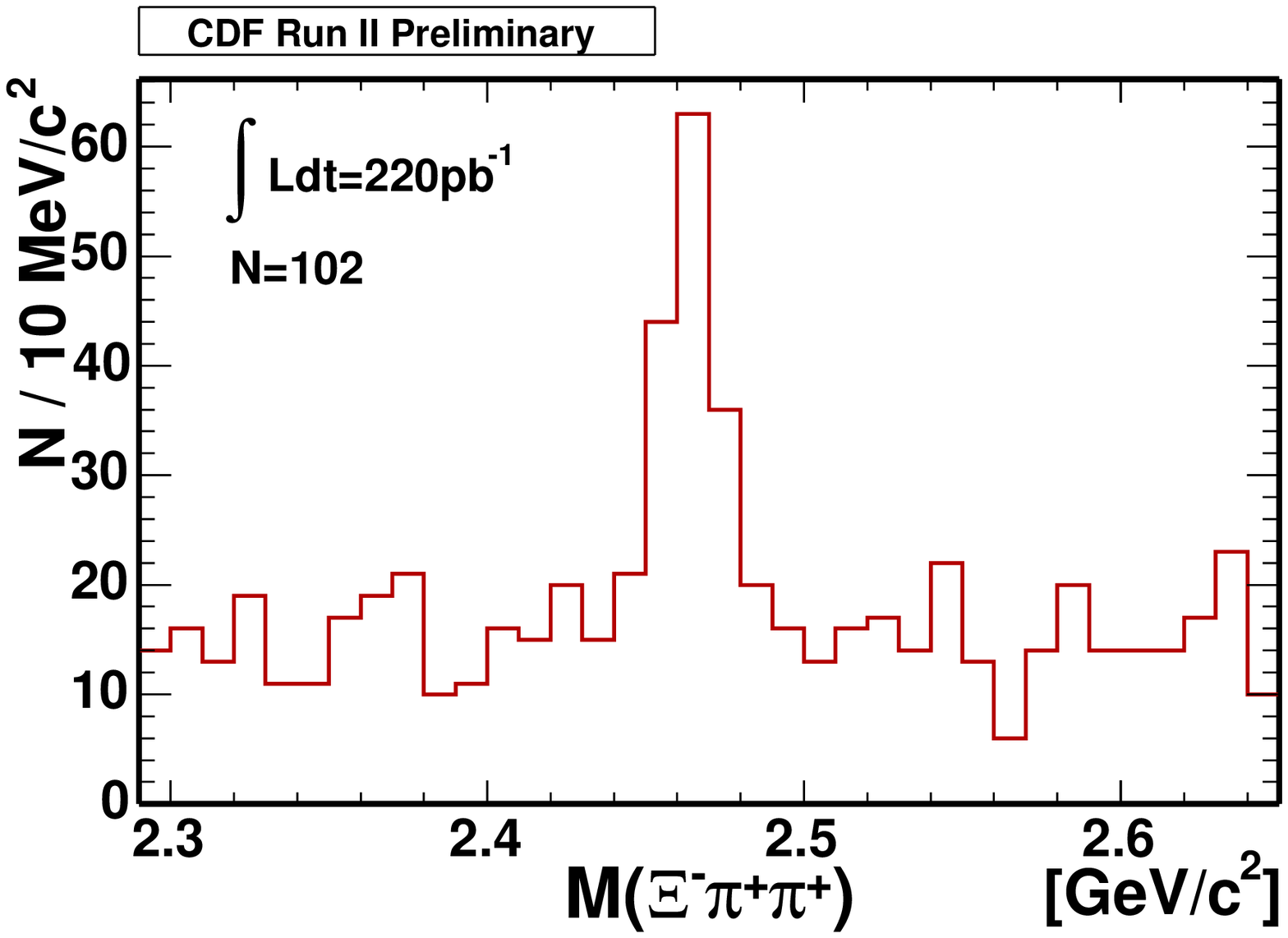,width=5.5cm}\\[-4mm]
\hspace*{1mm}\raisebox{-1mm}{\hspace*{2mm}(c)\hspace*{60mm}(d)}
\\[-1mm] 
\caption[*]{ Hyperon mass spectra observed by CDF.
In each case the charge-conjugate combination is included,
and in the first two plots the wrong-sign combination is shown for
comparison.
(a) $\Lambda\pi^-$, 
(b) $\Lambda K^-$, (c) $\Xi^-\pi^+$, (d) $\Xi^-\pi^+\pi^+$.\protect\\[-6mm]}
\end{figure}

In baryon physics, CDF have demonstrated the observation of a variety
of hyperon states, some of which contain charm as well as strangeness.
The method is based first on the reconstruction of $\Lambda\to p\pi^-$
decays; the cascade baryon $\Xi^-$ may then be reconstructed in its
$\Lambda\pi^-$ decay mode, the CDF tracking algorithm having been modified
to allow for the small kink in the observed track when the $\Xi$
decays into the $\pi$.  An extremely clean signal is observed (fig.\
2(a)).  The $\Omega^-$ particle is also prominently visible (fig.\
2(b)).

The $\Xi^-\pi^+$ and $\Xi^-\pi^+\pi^+$ spectra now show clear peaks
corresponding to the $\Xi_c(2470)$ in its neutral and singly-charged
states respectively (figs.\ 2(c,d)).  This is the first observation of these states
in $p\bar p$ collisions.

\begin{figure}[!t]
\epsfig{file=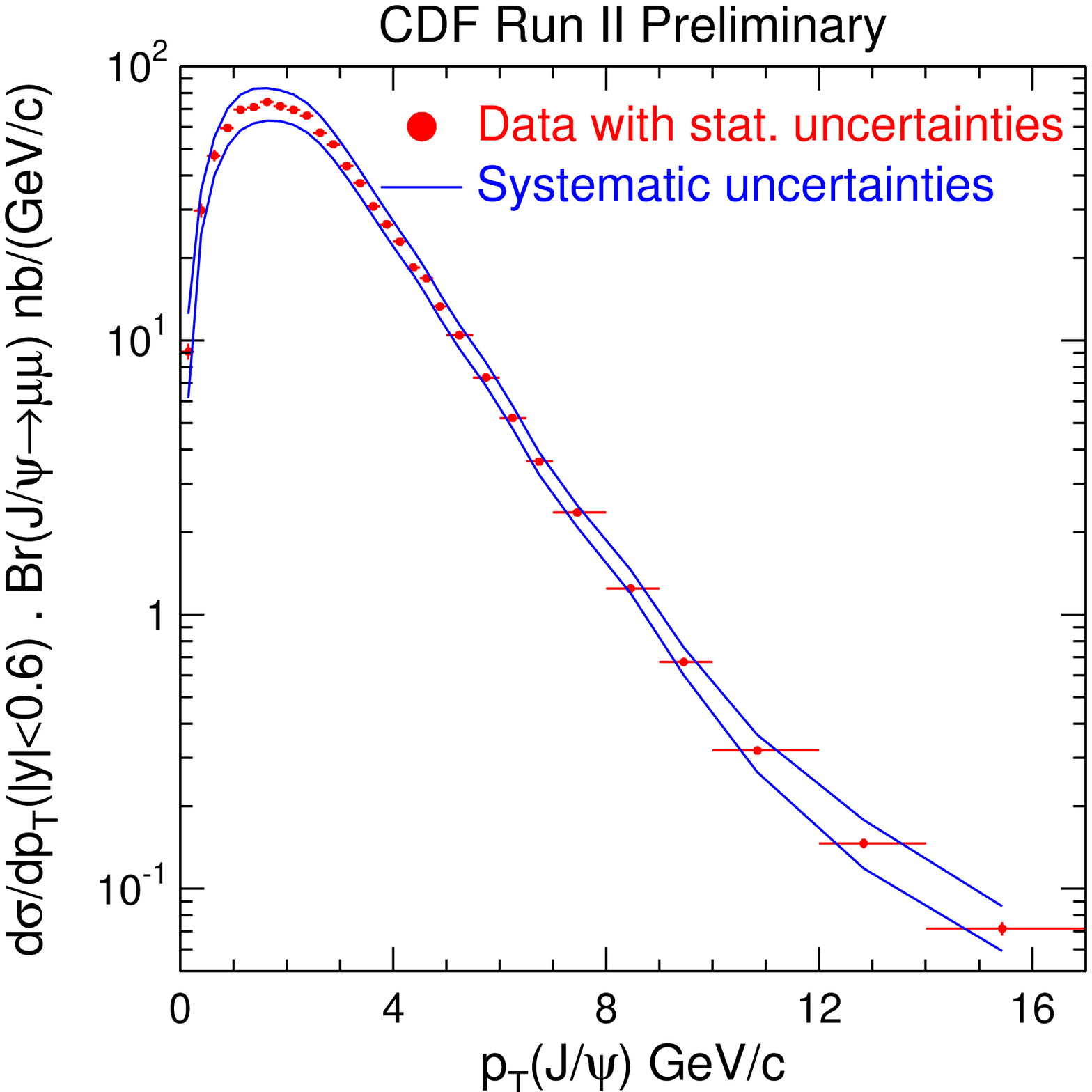,width=6.4cm}\hspace*{5mm}
\raisebox{-1mm}{\epsfig{file=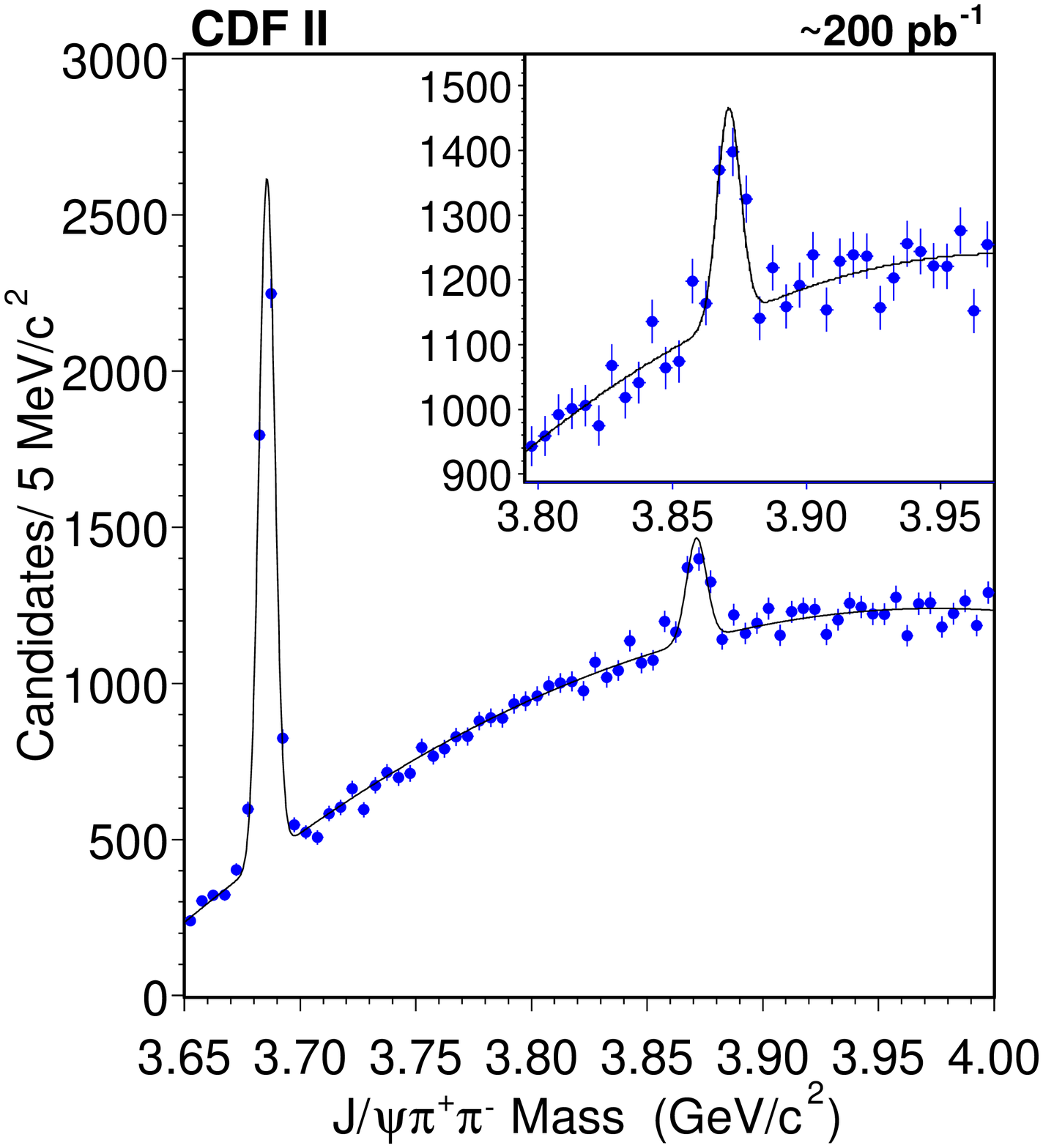,width=5.5cm}}
\\[-3mm]
\hspace*{1mm}\raisebox{-2mm}{\hspace*{0mm}(a)\hspace*{65mm}(b)}\\[0mm]
\caption{
(a) $J/\psi$ differential section observed by CDF as a function of $p_T$, 
(b) $J/\psi\pi^+\pi^-$ mass spectrum, showing peaks corresponding to the 
$J/\psi(4S)$ and $X(3872)$ states.} 
\end{figure}

\section{Charmonium  production}
The inclusive production of the $J/\psi$ and other $c\bar c$ states
was extensively studied at CDF I.  The process is evaluated theoretically
as a perturbative QCD calculation of the quark-antiquark state, together with 
a modelling of the subsequent hadronisation into the  observed $J/\psi$.
Both colour-singlet and colour-octet $Q\bar Q$ states may in principle be 
produced, and it is of interest to establish whether both mechanisms
are required to account for the observations.  On the 
basis of the $p_T$ distributions,
the Run I data demonstrated the need for both contributions. 
However the apparatus was then able to trigger only on  $J/\psi$
$p_T$ values above 4 GeV/c, restricting the full comparison with theory.
Using a new dimuon trigger, CDF is now able to record central  
 $J/\psi\to\mu^+\mu^-$ production down to $p_T$=0.  Preliminary results
are shown in fig.\ 3(a).  The measurable cross section is increased  by
a factor of approximately 14 compared to Run I.

In the summer of 2003 the BELLE experiment at KEK announced the
observation of a new charmonium state at 3872 MeV/c$^2$, decaying into
$J/\psi\pi^+\pi^-$.  The physical nature of this state is unclear,
since there is no very natural interpretation:  a heavy $^3D_2$ state
has been suggested, or possibly a $D^0{\bar D}^{*0}$ ``molecule''.
CDF have confirmed this discovery \cite{cdf3}. $J/\psi$ signals at
$p_T>4$ GeV/c were identified in the $\mu^+\mu^-$ decay mode, and
combined with charged pions with $p_T>0.4$ GeV/c.  A clear peak at 3872
MeV/c$^2$ is seen, the $\psi'(4S)$ peak being very clean and serving as a
control signal.  A further cut at $\pi^+\pi^-$ masses above 500 MeV/c$^2$
enhances the signal, as also found by BELLE.  This is shown in fig.\
3(b). The width is consistent with the resolution of the apparatus;
the fitted mass is 3871$\pm$0.7$\pm$0.4 MeV/c$^2$, in good agreement with
BELLE's value of 3871.7$\pm$.6 MeV/c$^2$.  

Subsequently CDF have shown that their signal is partly prompt and 
partly arising from long-lived decays, presumably of $B$ hadrons as must be the case in BELLE.
The long-lived fraction is found to be 16$\pm$4.9$\pm$2 \% (preliminary).

\section{Beauty production} 

Both CDF and D0 published cross sections for inclusive $B$ hadron
production in Run I.  The results from the two experiments were in
agreement, but lay significantly above the existing QCD predictions,
making an experimental confirmation in Run II an urgent necessity.
From the CDF $J/\psi$ cross sections described above, $B$ cross
sections have been extracted using an unfolding technique which
incorporates information from the SVT to identify the displaced decay
vertex of the $B$ state.  The fraction of $J/\psi$ arising from
initially produced $b$ quarks is approximately 10\% at $p_T$ values
around 2 GeV/c, rising to 50\% at 20 GeV/c (fig.\ 4(a)).  The
resulting differential cross sections agree well with those of Run I,
indicated in fig.\ 4(b) as the points with larger error bars.  There is also
agreement with the latest NLO QCD calculations represented by  
the FONLL predictions of Cacciari et al.\cite{cacciari2}, which are
plotted with their uncertainty. A similarly good agreement is
obtained with the MC@NLO Monte Carlo.  Both data and  theory are thus
significantly improved.  The areas of theory improvement lie especially in the 
proton PDF's that are employed, and in the description of the $b$
quark hadronisation.

\begin{figure}[!t]
\epsfig{file=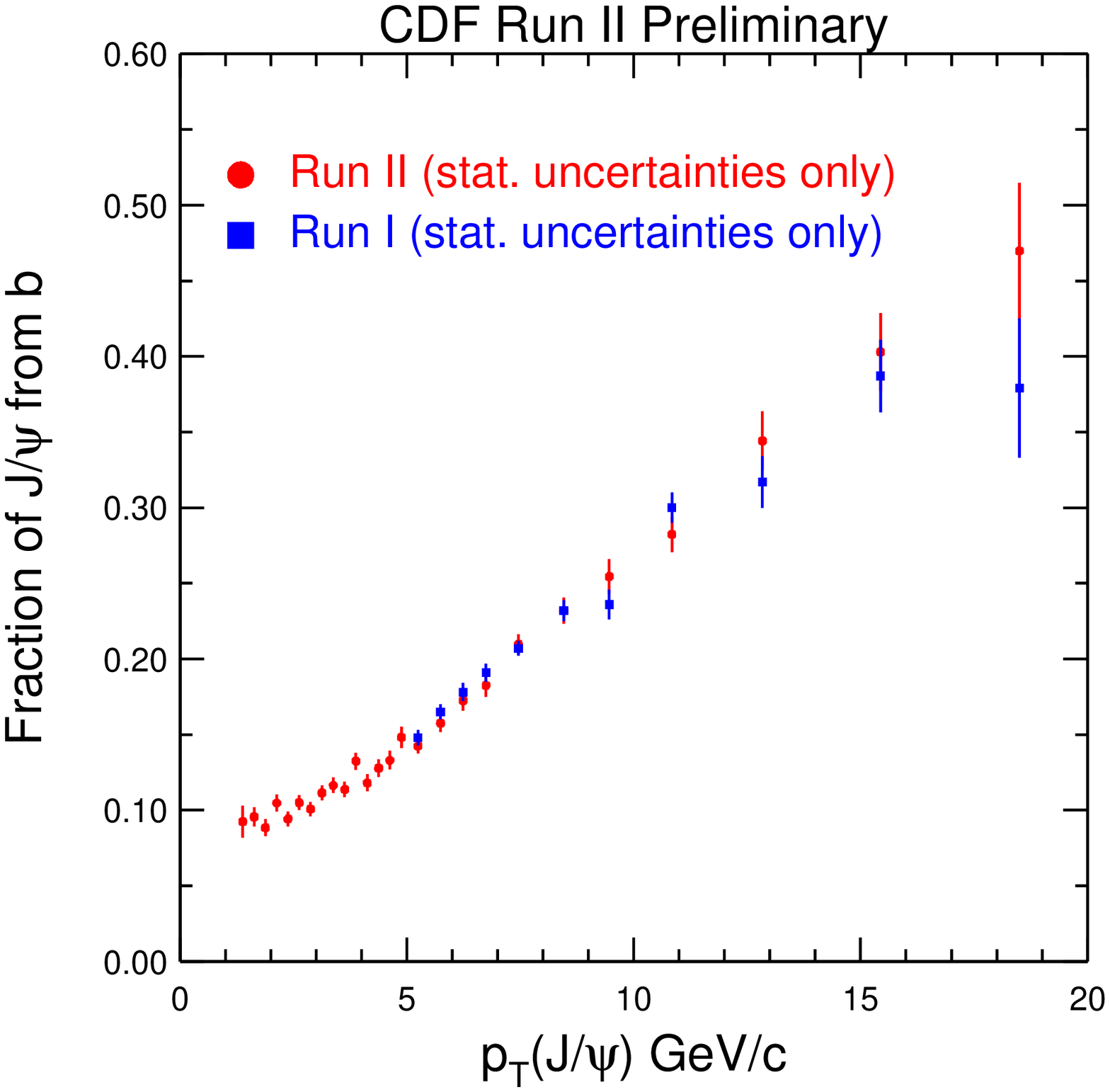,width=6.4cm}\hspace*{1mm}
\epsfig{file=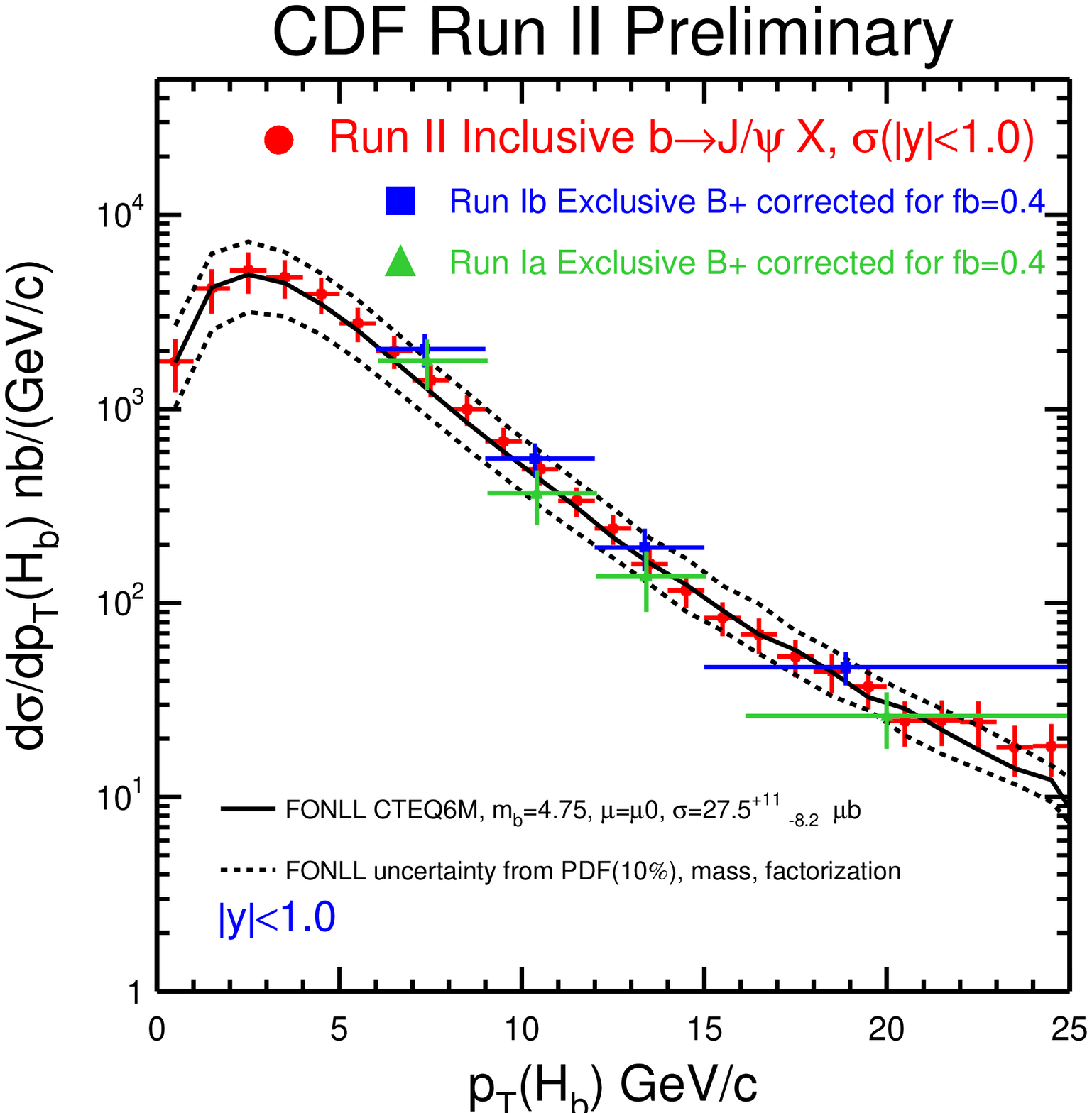,width=6.5cm}\\[-4mm]
\hspace*{1mm}\raisebox{-1mm}{\hspace*{0mm}(a)\hspace*{63mm}(b)}\\[-2mm] 
\caption{
(a) Fraction of inclusive
 $J/\psi$ which arise from the decay of a $B$ hadron,  
as a function of $p_T$,
(b) Differential cross section for inclusive $B$ hadron production
as a function of $p_T$.} 
\end{figure}

It is also found that PYTHIA can give a good description of the data,
provided that the flavour creation, flavour excitation and fragmentation
contributions are properly accounted for (fig.\ 5(a)).  The three
contributions are separately evaluated using PYTHIA 6.115 with the
CTEQ3L proton PDF's and a hard scattering $p_T$ threshold of 3 GeV/c.
A further study of this topic has been performed using Run I data on $B\bar
B$ correlations, The three contributions are plotted separately in a
comparison of the azimuthal separation $\Delta\phi $ of the $B\bar B$
particles in fig.\ 5(b), where the initial-state gluon radiation
generated by PYTHIA has also been optimally tuned.
The flavour creation component is strongly back-to-back peaked, 
the excitation component less so, while the gluon splitting
component is fairly flat in $\Delta\phi$.  The proportions are 
fixed at their predicted values, but the overall normalisation
is allowed to float.
A plausible fit is obtained; however it is not perfect and more
detailed studies will be valuable.

\section*{Summary}
In Run II of the Tevatron, the CDF Collaboration has expanded its
already flourishing programme of heavy flavour physics.  This enables
the latest theoretical models of heavy quark production to be tested
from a number of viewpoints.  Work is of course ongoing, but the improved
agreement between NLO QCD and the data already represents a triumph
both for theory and for experiment.

\begin{figure}[!t]
\epsfig{file=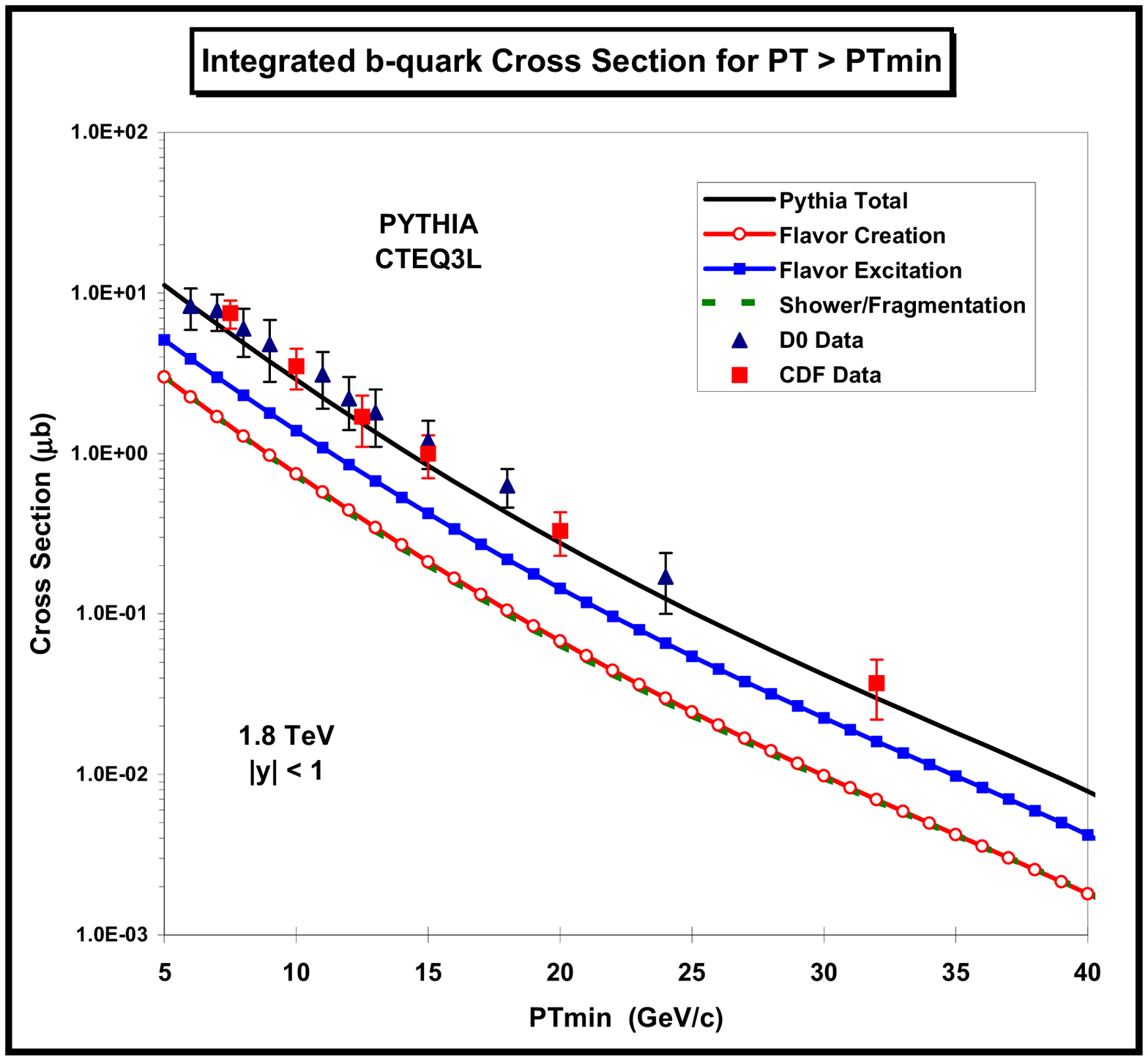,width=6cm}\hspace*{4mm}
\raisebox{-5mm}{\epsfig{file=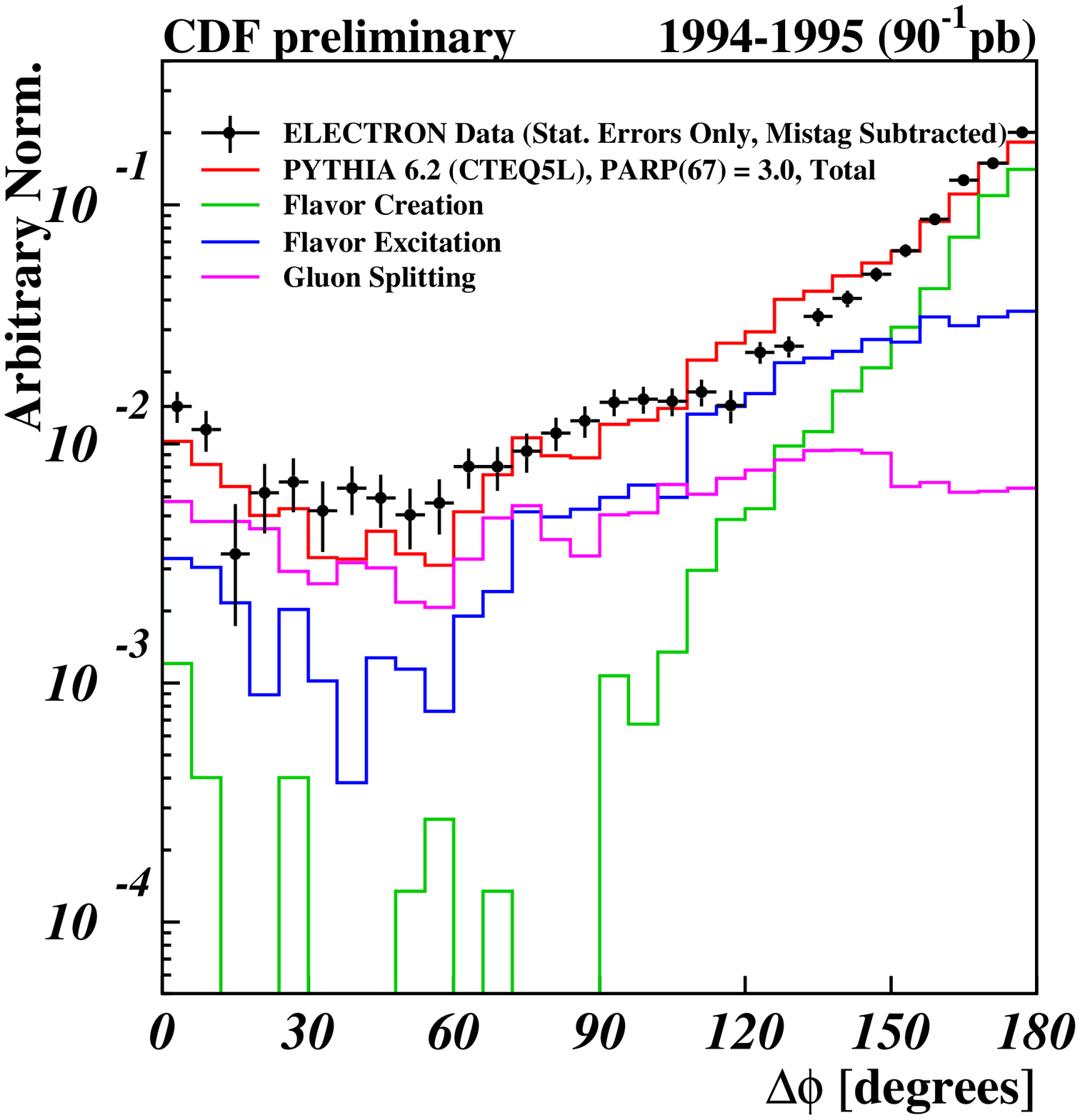,width=6.5cm}}
\\[-4mm]
\hspace*{1mm}\raisebox{-1mm}{\hspace*{0mm}(a)\hspace*{62mm}(b)}\\[-2mm]
\caption{
(a)
CDF and D0 Run I unfolded  cross sections for inclusive b-quark production
vs. $p_T$. The lower curve is the (coincident) contributions from flavour
creation and from fragmentation; the central curve is flavour creation and
the top curve is the sum.  All are evaluated using PYTHIA.   
(b) Azimuthal angle difference $\Delta\phi$ between $B$ hadrons having
$p_T$ values of $>$ 14 GeV/c (with an electron tag) and $>$ 7 GeV/c.
The upper histogram is the total predicted by PYTHIA from the three
component heavy flavour production mechanisms, plotted as the lower histograms.}
\end{figure}

\end{document}